\documentclass[%
 reprint,
 superscriptaddress, 
 amsmath,amssymb,
 prl,                
]{revtex4-2}

\usepackage{graphicx}
\usepackage{dcolumn}
\usepackage{bm}

\begin{document}

\title{Orientation engineering is a universal strategy for
  ferroelectric trans-switching}

\author{Cameron A.M. Scott}
 \affiliation{Smart Materials Unit, Luxembourg Institute of Science and Technology (LIST), Esch-sur-Alzette, Luxembourg}

\author{Jorge {\'I}{\~n}iguez-Gonz{\'a}lez}
\affiliation{Smart Materials Unit, Luxembourg Institute of Science and Technology (LIST), Esch-sur-Alzette, Luxembourg}
\affiliation{Department of Physics and Materials Science, University of Luxembourg, Belvaux, Luxembourg}

\begin{abstract}
About two decades ago, ``strain engineering'' emerged as a powerful
strategy to tune epitaxial thin films. Here we contend that
``orientation engineering'' -- that is, controlling the film's growth
direction -- constitutes a complementary design route that opens
unprecedented opportunities and deserves greater
attention. Specifically, we show that orientation engineering can
endow standard ferroelectric compounds with exotic and useful
behaviors that may be difficult or impossible to obtain
otherwise. Critically, we predict it readily enables
``trans-switching,'' where polarization components switch
perpendicularly to the applied electric field, paving the way for
novel 3D device architectures. We also show that, by controlling the
electric screening provided by electrodes in contact with the film,
ferrielectric behavior can be obtained. Notably, one can even design
an operational mode where the switching is, in effect, fully
transversal to the applied field. Supported by phenomenological
calculations for representative compounds LiNbO$_3$ and BaTiO$_3$, we
thus propose orientation engineering as a powerful and general method
to unlock unprecedented functionalities in common ferroelectrics.
\end{abstract}

\maketitle

\begin{figure}
    \centering
    \includegraphics[width=0.8\columnwidth]{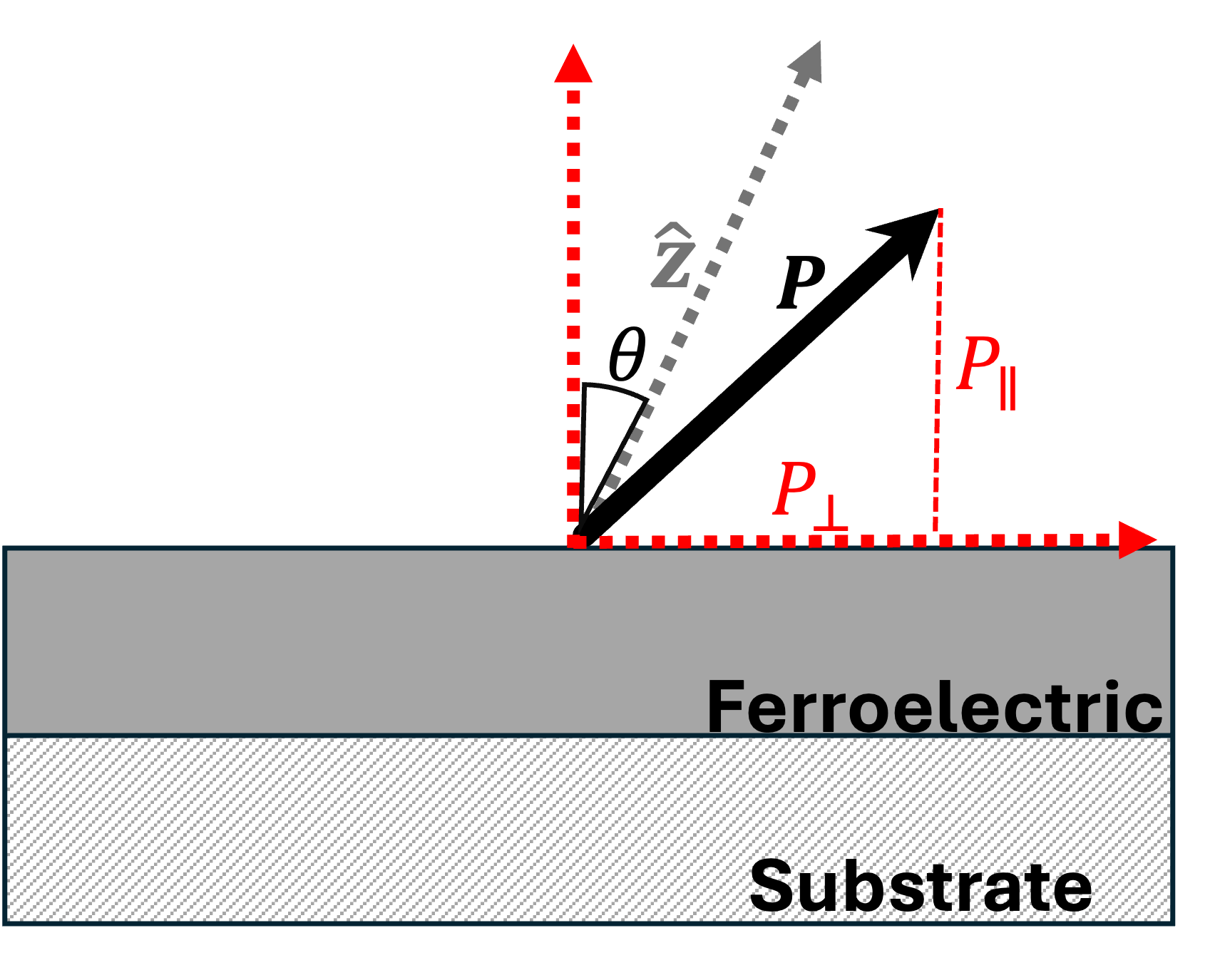}
    \caption{Schematic of the orientation-engineered films considered
      in this work. A principal direction of the crystalline lattice
      ($\hat{z}$) forms an angle ($\theta$) with the vertical growth
      direction. For arbitrary $\theta$, the polarization
      ($\mathbf{P}$) can be projected onto both out-of-plane
      ($P_\parallel$) and in-plane ($P_\perp$)
      components. Improper-like couplings between $P_{\parallel}$ and
      $P_{\perp}$ naturally occur for $\theta\neq 0$, leading to
      trans-switching.}
    \label{fig:intro}
\end{figure}

Epitaxial thin films are elastically constrained by the substrate on
which they are grown. This results in an epitaxial strain that enables
the well-known and much-used ``strain engineering'' of their
properties~\cite{schlom2007strain,schlom2014elastic}. Further, in
order to adapt to the epitaxial constraint, or to the features of the
substrate's surface (cut, morphology), films may adopt a particular
crystalline orientation. Mastering this aspect of film growth is
sometimes called ``orientation engineering''~\cite{sando2022strain},
but it remains largely unutilized. Here we discuss how orientation
impacts the properties of ferroelectric thin films, an important
family of functional materials. Our theoretical work reveals unique
effects that are largely driven by symmetry and can thus be expected
to be universal. They suggest that orientation engineering has the
potential to become a major research direction for the development of
novel nanodevices.

\begin{figure*}
    \centering
    \includegraphics[width=\textwidth]{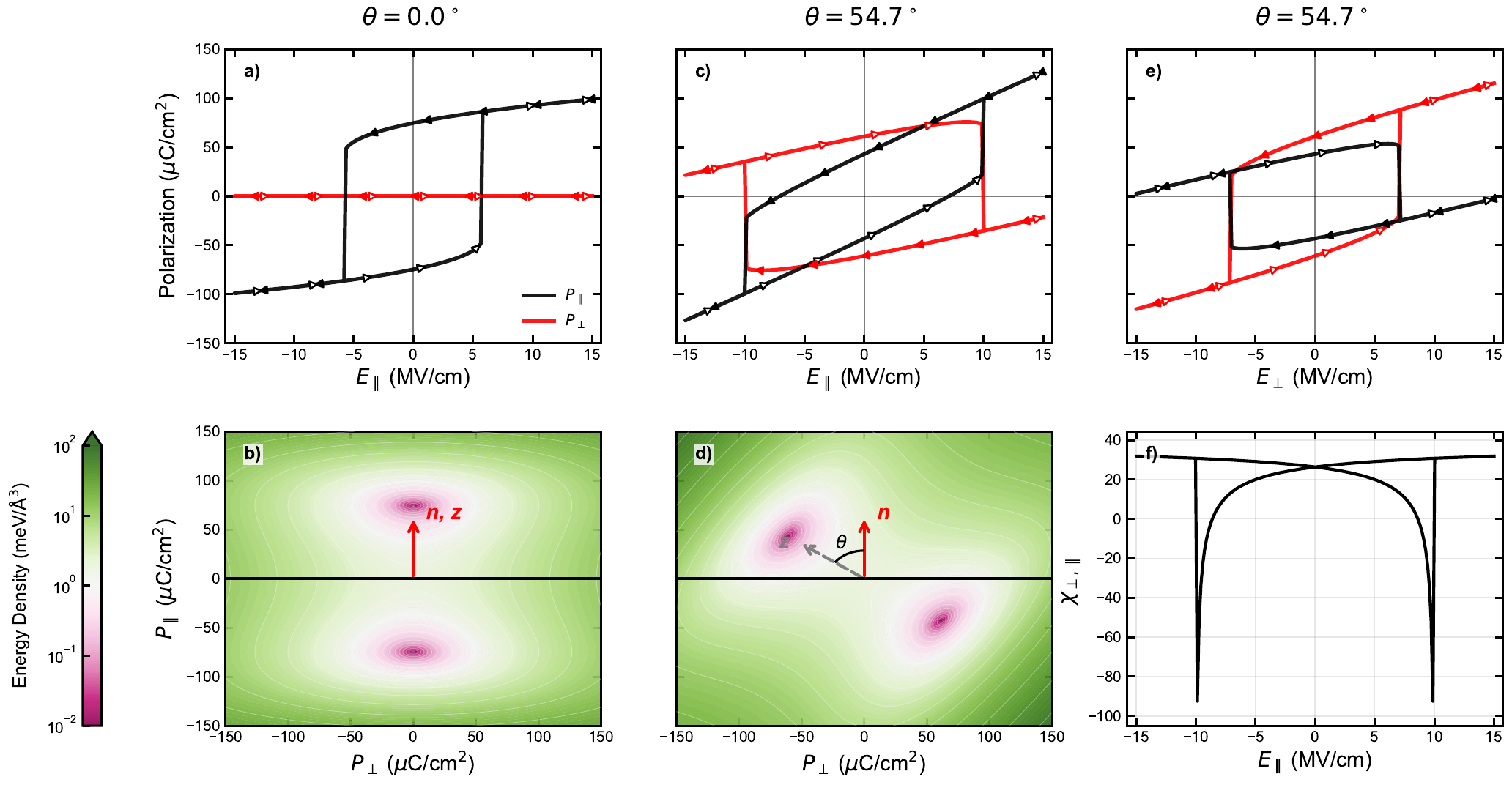}
    \caption{Trans-switching in LiNbO$_3$ films. Panels~a) and b)
      show, respectively, the hysteresis curve and zero-field free
      energy landscape a film with $\theta=0^\circ$ between the polar
      axis and the growth direction. A vertical electric field
      $E_\parallel$ leads to the standard hysteresis loop of the
      out-of-plane polarization $P_\parallel$. Panels~c) and d)
      correspond to the case with $\theta=54.7^\circ$. The non-zero
      angle imbues the equilibrium polarization with both in-plane and
      out-of-plane components that switch simultaneously under
      $E_\parallel$. Panel~e) shows the reciprocal effect of switching
      $P_\parallel$ by applying an in-plane field $E_\perp$ to the
      film with $\theta=54.7^\circ$. Panel f) shows the crossed
      dielectric susceptibility, $\epsilon_{0}\chi_{\perp,\parallel} =
      \partial P_{\perp}/\partial E_{\parallel}$, corresponding to the
      loop in panel~c).}
    \label{fig:LNO}
\end{figure*}

Let us focus on the question of ``trans-switching'' in ferroelectrics,
that is, the situation in which an applied electric field reverses a
polarization component orthogonal to it. Ferroelectric trans-switching
has been proposed as a mechanism enabling novel transcapacitive
elements that produce an output charge at one terminal in response to
a voltage applied across two other terminals, the capacitive analogue
of the transistor~\cite{mathuriya2026solid,di2009circuit}. These and
similar effects could be leveraged in novel devices, e.g. in 3D
architectures potentially enabling greater integration of electronic
and computing components~\cite{kim2016neurocube}.

Ferroelectric trans-switching may seem impossible at first sight, on
account of the absence of off-diagonal couplings in the electrostatic
energy $-\mathbf{E}\cdot\mathbf{P}$. Nevertheless, it has recently
been shown that this difficulty can be circumvented in a layered oxide
material by leveraging cross-couplings involving multiple lattice
distortions~\cite{gupta2026perpendicular}. Yet, however interesting,
it seems unlikely such complex mechanisms may be of widespread use.

Let us now see how trans-switching can be obtained in regular
ferroelectrics through orientation engineering. We discuss first the
case of a uniaxial ferroelectric with a single polar axis. Suppose
that $\hat{z}$ defines a principal direction of the crystalline
lattice, e.g. the polar axis itself. As shown in
Figure~\ref{fig:intro}, suppose we prepare our film so that $\hat{z}$
makes an angle $\theta$ with the vertical or growth direction, which
we make coincide with the surface normal for simplicity. An arbitrary
polarization state, characterized by $\bf{P}$, can be described in
terms of components parallel ($P_{\parallel}$) and perpendicular
($P_{\perp}$) to the growth direction. Typically, one acts on such a
film by applying an electric field along the vertical, thus affecting
$P_{\parallel}$ through the $-\mathbf{E}\cdot\mathbf{P}$ coupling
which simplifies to $-E_{\parallel}P_{\parallel}$.

\begin{figure*}
    \centering
    \includegraphics[width=\textwidth]{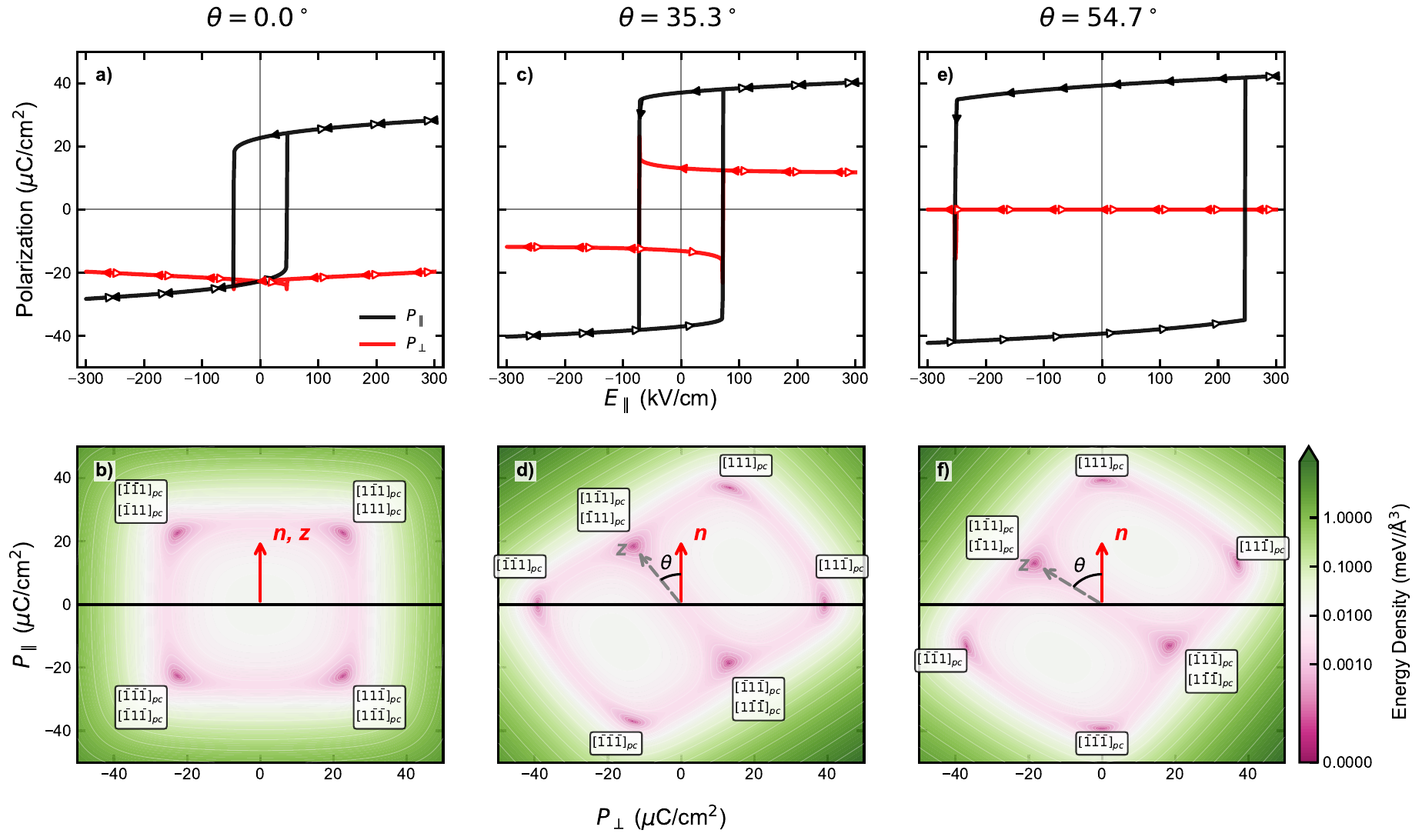}
    \caption{Trans-switching in rhombohedral BaTiO$_3$ at
      150~K. Panels a) and b) show, respectively, the hysteresis loops
      and zero-field energy landscape for a perfect alignment of
      the $[001]_{pc}$ pseudocubic axis with the growth direction
      ($\theta = 0^{\circ}$). We observe a 71$^\circ$ polarization
      rotation that only involves $P_{\parallel}$. Panels c-f) show, respectively, the corresponding
      results for $\theta=35.3^\circ$ ($[112]_{pc}$ orientation) and
      $\theta = 54.7^\circ$ ($[111]_{pc}$ orientation). Only one
      minimum has the maximum $P_\parallel$ component, and an electric
      field $E_\parallel$ drives the switching to the symmetrically
      opposite minimum, reversing all non-zero components and
      achieving trans-switching. The free-energy maps in the
      $(P_{\parallel},P_{\perp})$ plane show the result of a
      minimization along the second in-plane direction (orthogonal to
      both $P_{\parallel}$ and $P_{\perp}$). A minimum in this
      $(P_{\parallel},P_{\perp})$ plane may correspond to one or two
      minima of the three-dimensional polarization. To better
      understand the free-energy landscape and the actual crystalline
      orientations corresponding to $P_{\parallel}$ and $P_{\perp}$,
      in the maps we indicate the identity of the free-energy minima
      in the pseudocubic crystalline basis (see text).}
    \label{fig:BTO}
\end{figure*}

The basic behavior of a uniaxial ferroelectric can be captured by a
fourth-order Landau model of the form~\cite{toledano1987landau}
\begin{equation}
\label{eq:uniaxial}
 \mathcal{F}(P)=\alpha(T) P_z^2 + \beta P_z^4 + \gamma(P_x^2+P_y^2),
\end{equation}
where $\alpha$ is negative and temperature dependent, while $\beta$
and $\gamma$ are positive constants. Here, ${\bf P} =
(P_{x},P_{y},P_{z})$ is the polarization expressed in an orthogonal
basis where the $z$ is defined by $\hat{z}$ in Fig.~\ref{fig:intro}
and coincides with the polar axis. For simplicity, let us assume
$P_{y} = 0$ and work with a two-dimensional polarization ${\bf P} =
(P_{x},P_{z})$ that can be fully described by the film-adapted basis
${\bf P} = (P_{\perp},P_{\parallel})$ in Fig.~\ref{fig:intro}, a
generalization to three dimensions being trivial. We have
\begin{equation}
 P_{z}={\bf P} \cdot {\hat z} = P_{\parallel}\cos{\theta} +
 P_{\perp}\sin{\theta},
\end{equation}
and $P_{x}^{2} = P_{\parallel}^{2} + P_{\perp}^{2} - P_{z}^{2}$. We
can substitute these back into Eq.~(\ref{eq:uniaxial}) to obtain
\begin{equation}
\label{eq:free_energy}
\begin{split}
     \mathcal{F}(P_\perp,P_\parallel) &= \alpha_\parallel
     P_\parallel^2 + \alpha_\perp P_\perp^2 + fP_\parallel P_\perp
     \\&+ f'P_\parallel^3 P_\perp + f''P_\parallel P_\perp^3 \\&+
     \beta_\parallel P_\parallel^4 + \beta_\perp P_\perp^4 +
     gP_\parallel^2 P_\perp^2,
\end{split}
\end{equation}
where all the coefficients depend on $\theta$. For example,
$\alpha_\parallel = \alpha\cos^2{\theta} + \gamma\sin^2{\theta}$,
$\alpha_\perp = \alpha\sin^2{\theta} + \gamma\cos^2{\theta}$,
$f=2(\alpha-\gamma)\cos{\theta}\sin{\theta}$ and
$f'=4\beta\cos^3{\theta}\sin{\theta}$.

Notably, we observe the appearance of terms coupling $P_{\perp}$ and
$P_{\parallel}$, involving odd-powers of the individual polarization
components. The corresponding coupling coefficients ($f$, $f'$ and
$f'')$ are non-zero as long as $\theta \neq \frac{n\pi}{2}$ where $n$
is an integer. Couplings of this kind, irrespective of the sign of the
coefficient, promote the simultaneous occurrence, and the simultaneous
switching, of both components to keep the energy minimised. They are
analogous to those that drive polar distortions in improper
ferroelectrics~\cite{levanyuk1974improper,benedek2011hybrid,senn2018group}.

We now study the effects of these improper-like terms in
LiNbO$_{3}$~\cite{megaw1968note,bartasyte2017toward}, whose high
crystalline anisotropy results in a near perfect manifestation of the
uniaxial ferroelectric studied in Eq.~(\ref{eq:uniaxial}). We
construct the free-energy landscape using a Landau model available in
the literature and derived from experimental
data~\cite{scrymgeour2005phenomenological,chen2007appendix}. To study
the effect of an out-of-plane electric field, we add a
$-E_{\parallel}P_{\parallel}$ term to the model. (Eventually, we also
consider an in-plane field through a $-E_{\perp}P_{\perp}$ term.) To
simulate switching from a given state we proceed as follows: we track
all the local minima in the free energy as a function of the changing
electric field and, when the currently occupied minimum is destroyed,
consider that the system switches to the global minimum. (One may
alternatively consider that the system falls into the minimum within
its basin of attraction. In the SI we show that the qualitative
behavior remains unchanged.) We run our simulations for LiNbO$_{3}$ at
300~K and assume ideal short-circuit electric boundary conditions, so
depolarizaing fields are zero.

Figure~\ref{fig:LNO}a shows our results for $\theta=0^\circ$ and a
vertical field applied. Here we obtain a standard hysteresis loop,
whereby the vertical polarization $P_{\parallel}$ switches between the
two energetically degenerate minima in the free energy landscape
depicted in Fig.~\ref{fig:LNO}b, while $P_{\perp} = 0$ throughout.

This contrasts with the case in Fig.~\ref{fig:LNO}c, obtained by
choosing $\theta = 54.7^\circ$. We still apply a vertical field
$E_{\parallel}$, but here observe switching in both $P_{\parallel}$
and $P_{\perp}$, and so trans-switching is achieved. Note that this
growth direction leads to the rotated free-energy landscape in
Fig~\ref{fig:LNO}d, imbuing the remanent polarization with in-plane
and out-of-plane components. Since there are only two minima in a
uniaxial ferroelectric like LiNbO$_{3}$, field-induced switching from
one to the other must necessarily involve a reversal of both
components of the polarization. This makes perpendicular switching a
completely general effect in orientation-engineered uniaxial
ferroelectrics. In our simulations of LiNbO$_{3}$, we observe it for
any $\theta$ different from $0^{\circ}$ or $90^{\circ}$.

If we now apply an in-plane field $E_\perp$ to the film with
$\theta=54.7^\circ$, we observe trans-switching of $P_\parallel$
(Fig.~\ref{fig:LNO}e). This illustrates that orientation-engineered
uniaxial ferroelectrics possess the capacity for robust and reciprocal
trans-switching. Note that the coercive fields for trans-switching are
of the same magnitude as those observed in the regular hysteresis loop
($\theta=0^{\circ}$, Fig~\ref{fig:LNO}a). Finally, in
Fig.~\ref{fig:LNO}f we show the off-diagonal dielectric susceptibility
corresponding to the loop in Fig.~\ref{fig:LNO}c. Large, strongly
field-dependent transversal responses emerge from the same
improper-like couplings that lead to trans-switching. Hence,
orientation-engineered uniaxial ferroelectrics would be ideally suited
for use as highly-tunable transcapacitors.

How does this idea work in triaxial ferroelectrics where the
free-energy landscape is much richer? To address this, we study
perovskite oxide BaTiO$_{3}$, a prototype triaxial compound. As
detailed in the SI, one can consider an appropriate fourth-order
Landau model and conduct an analysis analogous to the one described
above for uniaxial materials. The main conclusion is that, in this
case too, improper-like couplings appear as the growth direction
differs from the highest-symmetry crystalline axes. Interestingly, the
bilinear cross-coupling is never present (i.e., $f=0$ when $\alpha =
\gamma$ in Eq.~(\ref{eq:uniaxial})), but quartic interactions
essentially identical to those quantified by $f'$ and $f''$ in
Eq.~(\ref{eq:uniaxial}) are active for nearly all orientations.

To make this discussion specific, we study orientation-engineered
BaTiO$_3$ using an experimentally derived Landau
model~\cite{li2005phenomenological,chen2007appendix}. We set $T=150$~K
to focus on the case where we have eight rhombohedral minima of the
free energy, and assume short-circuit electric boundary
conditions. Figure~\ref{fig:BTO}a shows our result for a film where
the growtn direction coincides with the pseudocubic axis $[001]_{pc}$,
which we identify with $\hat{z}$. This is the case with $\theta =
0^{\circ}$, for which we find a regular switching behavior involving
only $P_{\parallel}$. Note that this is so even though, at variance
with the case of LiNbO$_{3}$, here the $\theta = 0^{\circ}$
orientation involves a spontaneous polarization with both out-of-plane
and in-plane components. Yet, as can be appreciated in
Fig.~\ref{fig:BTO}b, the free-energy landscape is highly symmetric and
the material can reverse $P_{\parallel}$ without needing to change the
in-plane polarization. Consistent with this, and as shown in the SI,
no improper-like couplings are allowed by symmetry for a
$[001]_{pc}$-oriented BaTiO$_{3}$ film.

Before discussing other growth directions, note that, to study a
triaxial ferroelectric like BaTiO$_{3}$, one must in principle keep
track of both in-plane polarization components. Nevertheless, our
observations concerning trans-switching do not depend on how the
in-plane polarization axes are chosen, which implies that, for the
current purposes and without loss of generality, we can focus on a
single $P_{\perp}$ direction and a single angle $\theta$ is sufficient
to characterize the film's orientation. The free-energy maps in
Fig.~\ref{fig:BTO} constitute a simplified representation whereby we
track the minimum free energy for given $(P_{\parallel},P_{\perp})$
coordinates, performing a minimization along the second (not shown)
in-plane direction. In particular, each minimum shown in
Fig.~\ref{fig:BTO}b corresponds to two minima of the free-energy
landscape; for example, the rhombohedral minima with polarizations
along $[111]_{pc}$ and $[1\bar{1}1]_{pc}$ share
$(P_{\parallel},P_{\perp})$ coordinate, which implies that, in this
case, $P_{\perp}$ lies along the $[100]_{pc}$ direction while the
second (not shown) in-plane direction is $[010]_{pc}$.

Figures~\ref{fig:BTO}c and \ref{fig:BTO}d show the case of a film
grown along the $[112]_{pc}$ direction ($\theta = 35.3^{\circ}$). The
spontaneous polarization presents both vertical and in-plane
components, and trans-switching occurs. The free-energy map is a
rotated version of that for $\theta=0^{\circ}$. While all the
equilibrium states are equivalent by symmetry (and degenerate) at zero
field, they differ in how their polarization splits between in-plane
and out-of-plane. In particular, the minima presenting the largest
$|P_{\parallel}|$ components (which correspond to $[111]_{pc}$ and
$[\bar{1}\bar{1}\bar{1}]_{pc}$) are the ones occurring in the
hysteresis loop of Fig.~\ref{fig:BTO}c, as the switching is driven by
a vertical field that selects them among competing variants. The
free-energy map shows how a transition between such minima involves a
reversal of all polarization components, reflecting the underlying
improper-like couplings.

Finally, Figs.~\ref{fig:BTO}g and \ref{fig:BTO}f display the case
where the film is oriented along $[111]_{pc}$ ($\theta =
54.7^{\circ}$), that is, a polar axis. Here we do have free-energy
minima whose polarization presents in-plane and out-of-plane
components. However, the states expected to dominate the switching
driven by a vertical field are those with the largest
$|P_{\parallel}|$, perfectly aligned with the growth direction in this
case. Hence, here we do not observe trans-switching under
$E_{\parallel}$ even if the free-energy contains improper-like
couplings. By contrast, an in-plane field would cause the film to
switch between, e.g., states $[11\bar{1}]_{pc}$ and
$[\bar{1}\bar{1}1]_{pc}$, which will lead to trans-switching. This
behavior is apparent from the free-energy map in Fig.~\ref{fig:BTO}f
and explicit results are provided in the SI.

\begin{figure*}
    \centering
    \includegraphics[width=\textwidth]{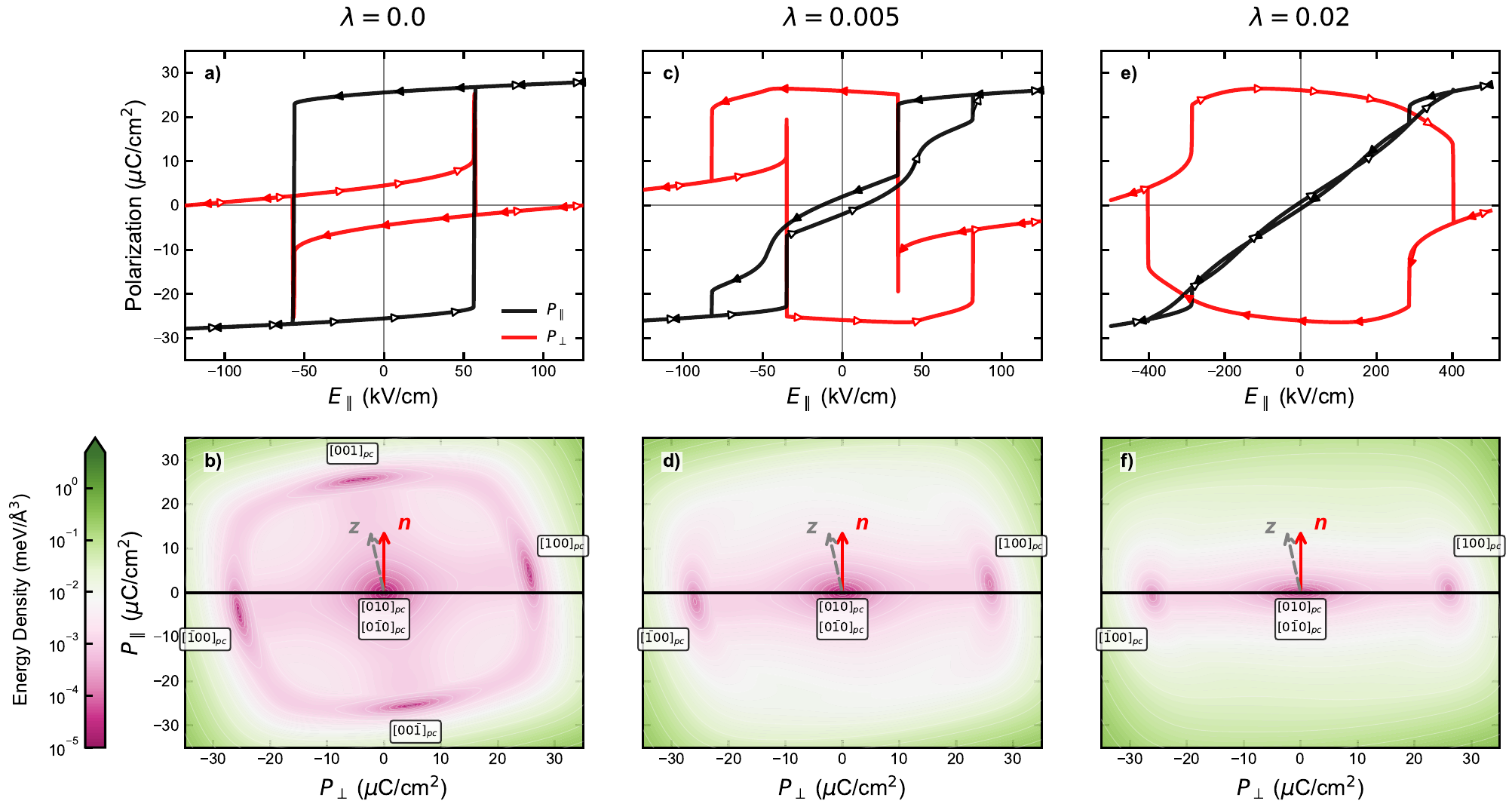}
    \caption{Effect of screening on tetragonal BaTiO$_3$ at 300~K with
      $\theta=10^{\circ}$ orientation. Panel~a) shows trans-switching
      for perfect short circuit boundary conditions ($\lambda=0$), as
      the vertical field $E_{\parallel}$ connects the two free-energy
      minima with maximum $|P_\parallel|$ in panel~b). Increasing
      $\lambda$ in Panels c) and d) removes these minima at zero field
      field (although they can still be induced by a finite
      $E_\parallel$), so the remanent equilibrium states display large
      $P_{\perp}$ and a relatively small $P_{\parallel}$. This results
      in a triple hysteresis loop. Further increasing $\lambda$
      removes the remanent $P_\parallel$ almost completely, leading to
      a situation with a nearly purely linear dielectric response for
      $P_{\parallel}$, and (trans-)switching restricted to
      $P_{\perp}$.}
    \label{fig:screening}
\end{figure*}

So far we have considered ideal short-circuit electric boundary
conditions, implying perfect electrostatic screening by the electrodes
in contact with the ferroelectric and no depolarizing
field~\cite{mehta1973depolarization}. This is a realistic scenario, as
one can routinely pole LiNbO$_3$ and BaTiO$_3$ capacitors to reach an
essentially monodomain state that displays a well-developed and stable
polarization. Nevertheless, it is known that, under imperfect
screening, the energy landscape can be further engineered and
optimized, enabling e.g. the stabilization of exotic electric textures
\cite{kornev2004ultrathin,kim2005polarization} or the emergence of
functional
properties~\cite{glazkova2014tailoring,aramberri22,mundy2022liberating,yin24}. Now
we explore whether imperfect electrode screening may allow us to
further tune the behavior of orientation-engineered ferroelectrics.

To address this, we incorporate the depolarizing effects in our model
though a term of the form $+\frac{\lambda}{2\epsilon_0}P_\parallel^2$,
where the dimensionless parameter $\lambda$ ranges from 0 (perfect
screening) to 1 (no screening, open-circuit boundary
conditions). (Instead of $\epsilon_{0}$, here one could use a
background permittivity $\epsilon_{b}$ to better adapt to the
definition of $\bf{P}$ in the employed Landau model. This is a
non-essential choice that merely affects the scale of the auxiliary
parameter $\lambda$.) We discuss the relevant and illustrative case of
room temperature BaTiO$_3$ with $\theta = 10^\circ$, a configuration
that could be realised using a miscut film grown on a vicinal
substrate~\cite{zhang2018surface}. Note that, at room temperature,
BaTiO$_{3}$ presents six tetragonal free-energy minima lying along
$\langle 001 \rangle_{pc}$ directions. Figure~\ref{fig:screening}
summarizes our main results.

Figures~\ref{fig:screening}a and \ref{fig:screening}b correspond to
$\lambda = 0$. We find that the film switches between the states with
the largest $|P_{\parallel}|$ (i.e., those associated to $[001]_{pc}$
and $[00\bar{1}]_{pc}$) and trans-switching of the small $P_{\perp}$
component occurs. This case is strongly reminiscent of trans-switching
in uniaxial LiNbO$_{3}$ (see Fig.~\ref{fig:LNO}c).

By contrast, Figs.~\ref{fig:screening}c and \ref{fig:screening}d show
the result for $\lambda = 0.005$. For imperfect screening, the
free-energy minima stop being degenerate, even at zero applied field,
those with a large in-plane polarization component being strongly
favored. Specifically, for this choice of $\lambda$, the states
corresponding to $[001]_{pc}$ and $[00\bar{1}]_{pc}$ stop being minima
of the free energy at zero field (hence, they are not labeled in
Fig.~\ref{fig:screening}d); yet, they reemerge when a sufficiently
large $E_{\parallel}$ is applied. The minima to the left and right of
Fig.~\ref{fig:screening}d (corresponding to the $[\pm 100]_{pc}$
directions) present a non-zero $P_{\parallel}$ component, despite the
imperfect screening; such small vertical polarization emerges from the
improper-like couplings, mainly of the form
$P_{\parallel}P_{\perp}^{3}$. Finally, the states corresponding to
$[0\mathord{\pm} 10]_{pc}$ do not present any vertical polarization component
for the chosen film orientation; they are the most stable states at
zero field (as they do not suffer from any depolarization), but become
dominated by the $[\mathord{\pm} 100]_{pc}$ states, and eventually by the
field-induced $[00\mathord{\pm} 1]_{pc}$ minima, as the applied
$|E_{\parallel}|$ grows. These observations explain the loop of
Fig.~\ref{fig:screening}c, which features a ferrielectric-like triple
hysteresis for
$P_{\parallel}$~\cite{maisonneuve1997ferrielectric,fu2020unveiling}. Simultaneously,
the $P_{\perp}$ component undergoes trans-switching that occurs in the
back-transformation, from $[00\mathord{\pm} 1]_{pc}$-like states to $[\mathord{\pm} 1
  00]_{pc}$-like states, upon removal of the applied
$E_{\parallel}$. (In the SI we show the evolution of the free-energy
maps with applied field, which helps visualize the mentioned
transformations.) This incredible behavior illustrates the unique
possibilities offered by imperfect screening combined with orientation
engineering.

We further reduce the quality of the screening and set
$\lambda=0.02$. The results in Figs.~\ref{fig:screening}e and
\ref{fig:screening}f show a behavior where the remanent vertical
polarization is all but fully suppressed, while the field-induced
stabilization of the $[00\mathord{\pm}1]_{pc}$ states (now occurring at
significantly higher fields) still enables the trans-switching of the
in-plane component. This illustrates the striking possibility of
confining the ferroic response to a direction perpendicular to the
applied field. It should be noted, though, that in the shown case all
the four states with a large in-plane polarization become essentially
degenerate at zero field, which might complicate a deterministic
trans-switching. We expect this issue could be resolved by choosing a
growth direction that further breaks the symmetry between the $[\mathord{\pm}
  100]_{pc}$ and $[0\mathord{\pm} 10]_{pc}$ directions, so that one pair
clearly dominates over the other.

Repeating the screening analysis for uniaxial LiNbO$_3$ at $\theta =
10^{\circ}$, we make the following observations. First, since in this
case we have only two minima of the free energy, we lack the
complexity necessary to obtain triple loops as that in
Fig.~\ref{fig:BTO}c. Second, in this case a finite remanent
polarization $P_\parallel$ and a trans-switching $P_\perp$ exist even
for $\lambda=1$; this is consistent with the well-known
hyperferroelectricity of LiNbO$_3$
\cite{garrity2014hyperferroelectrics} and suggests that
trans-switching should be observable even in films with an open
surface.

In conclusion, we have shown that orientation engineering can readily
enable complex behaviors in ferroelectric thin films, an observation
that we expect will be echoed for other properties and materials. In
particular, we have seen that, by choosing a growth direction of
relatively low crystalline symmetry, one can automatically induce
improper-like couplings between out-of-plane and in-plane polarization
components, which immediately leads to trans-switching in common
ferroelectrics like LiNbO$_{3}$ and BaTiO$_{3}$. Our results further
illustrate how this strategy can be enhanced by controlling the
quality of the electrostatic screening provided by the electrodes in
contact with the ferroelectric film; we predict that multiple loops,
and even an operational mode where the switching is effectively
constrained to the plane perpendicular to the applied electric field,
can thus be achieved. Critically, all our observations rely on
symmetry-driven effects that can be expected to be universal; further,
they can be understood in simple terms by considering free-energy maps
rotated to match the growth directions, which will facilitate the
prediction and engineering of further effects in the future. We also
note that the growth directions discussed here are already accessible
in thin films~\cite{angsten2017orientation,martin2016thin}, including
LiNbO$_3$~\cite{sando2018epitaxial,lee1996control} and organic
perovskites of interest for solar cell
applications~\cite{cho2016pure,li2024coherent,chen2021oriented}. Importantly,
all of our conclusions should also hold for the recently developed
free-standing membranes~\cite{ji2019freestanding}. For all these
reasons, we believe orientation engineering to be an intuitively
simple method for extending the range of ferroelectric thin film
functionality, especially towards the design of novel transcapacitive
elements. We look forward to a widespread exploration and further
development of these ideas.

\textit{Acknowledgements} - This work was supported by the Luxembourg
National Research Fund (FNR, grant INTER/ANR/24/18960894/TOPOTHERM)
and the French National Research Agency (ANR, grant ANR-24-CE30-5392),
within the framework of the ANR–FNR collaboration TOPOTHERM.

\bibliography{apssamp}

\end{document}